\documentclass{ifacconf}

\usepackage{graphicx}
\usepackage{natbib}
\usepackage{xcolor}
\usepackage{amsmath,amssymb}
\usepackage{mathtools}
\usepackage{algorithm}
\usepackage{algpseudocode}
\usepackage{tabularx}
\usepackage{xurl}

\newcommand{\R}{\ensuremath{\mathbb{R}}}

\newcommand{\diag}{\operatorname{diag}}

\def\T{\mathsf{T}}
\def\bmat{\begin{bmatrix}}
\def\emat{\end{bmatrix}}

\newcommand{\para}[1]{\left(#1\right)}		% parantheses ()
\begin{document}
\begin{frontmatter}

\title{Distributed Traffic State Estimation in V2X-Enabled Connected Vehicle Networks\thanksref{footnoteinfo}}

\thanks[footnoteinfo]{This work was supported by the Swedish Research Council's Distinguished Professor Grant, the Knut and Alice Wallenberg Foundation's Wallenberg Scholar Grant, and Digital Futures' Summer Research Internship Programme.}

\author[kth,groningen]{Vincent de Heij} 
\author[kth]{M. Umar B. Niazi} 
\author[groningen]{Saeed Ahmed}
\author[kth]{Karl H. Johansson}

\address[kth]{Division of Decision and Control Systems, Digital Futures, KTH Royal Institute of Technology, SE-100 44 Stockholm, Sweden (emails: vidh@kth.se; mubniazi@kth.se; kallej@kth.se)}
\address[groningen]{Jan C. Willems Center for Systems and Control, and the Engineering and Technology Institute Groningen, Faculty of Science and Engineering, University of Groningen, 9747 AG Groningen, The Netherlands (email: s.ahmed@rug.nl)}

\begin{abstract}
This paper presents a distributed traffic state estimation framework in which infrastructure sensors and connected vehicles act as autonomous, cooperative sensing nodes. These nodes share local traffic estimates with nearby nodes using Vehicle-to-Everything (V2X) communication. The proposed estimation algorithm uses a distributed Kalman filter tailored to a second-order macroscopic traffic flow model. To achieve global state awareness, the algorithm employs a consensus protocol to fuse heterogeneous spatiotemporal estimates from V2X neighbors and applies explicit projection steps to maintain physical consistency in density and flow estimates. The algorithm’s performance is validated through microscopic simulations of a highway segment experiencing transient congestion. Results demonstrate that the proposed distributed estimator accurately reconstructs nonlinear shockwave dynamics, even with sparse infrastructure sensors and intermittent vehicular network connectivity. Statistical analysis explores how different connected vehicle penetration rates affect estimation accuracy, revealing notable phase transitions in network observability.
\end{abstract}

\begin{keyword}
Distributed Traffic State Estimation;
Connected Vehicles;
V2X Communication;
Distributed Kalman Filter;
Aw-Rascle-Zhang Model.
\end{keyword}

\end{frontmatter}
%===============================================================================

\section{Introduction}
Intelligent transportation systems rely on a heterogeneous sensing architecture for traffic monitoring. 
This framework integrates data from fixed infrastructure sensors (e.g., induction loops, radars, cameras) and mobile probe data provided by connected vehicles (CVs) via vehicle-to-everything (V2X) communication. 
Infrastructure sensors, or roadside units (RSUs), provide accurate measurements of flow and occupancy, but are stationary and sparsely distributed. 
In contrast, CVs are mobile and equipped with advanced onboard sensors (e.g., GPS, cameras, and LiDARs) that provide a localized, microscopic view of the traffic stream in their immediate surroundings. 
Integrating these sources enables a comprehensive, multi-scale monitoring of network traffic conditions.

Using real-time measurements from both RSUs and CVs, traffic state estimation (TSE) methods reconstruct critical variables, such as traffic density and mean velocity, across the entire highway segment. 
Historically, TSE methods have relied on a centralized data-fusion architecture, in which all measurements are transmitted to a transportation management center (TMC) for global estimation.
This approach, however, is fundamentally non-scalable and ignores critical physical and communication constraints inherent to the V2X environment.
Firstly, V2X communication protocols (e.g., dedicated short-range communication (DSRC) and cellular V2X (C-V2X)) are constrained to a limited range, typically 300-500 meters \citep{ansari2021, zadobrischi2024, mendes2025}. 
This constraint creates an inherently sparse, localized communication topology that is not suitable for centralized estimation methods. 
Secondly, as the CV penetration rate soars\!\footnote{Statista 2025: \textit{Connected cars as a share of the total car parc in the EU}. [Available Online]}\!\!, the sheer volume of high-frequency sensor data generated by the collective fleet would require an excessive bandwidth, leading to prohibitive communication latency and a non-scalable computational burden if centralized \citep{arthurs2021}. 
This latency is not suitable for real-time control applications and also introduces an unacceptable single point of failure.

The core challenge, therefore, is not just the sparsity of data, but the need to process dense, localized, short-range data streams under strict latency requirements. 
This is particularly critical when dealing with dynamic, non-equilibrium traffic conditions. 
The literature on TSE has developed powerful model-based approaches, including first-order models (e.g., LWR~\citep{lighthill1955, richards1956} and CTM~\citep{daganzo1994}) and second-order models (e.g., ARZ \citep{aw2000, zhang2002}), with the latter being superior for capturing complex, non-equilibrium phenomena such as shockwaves and capacity drops. 
While many TSE methods \citep{Herrera2010, hinsbergen2012, yuan2012, bekiaris2016, fountoulakis2017, seo2017review, vishnoi2024} achieve high accuracy, they remain confined to the non-scalable centralized paradigm. 
On the other hand, a limited number of studies \citep{vivas2015, sun2014, sun2017, guo2017, zhang2020} have explored distributed TSE, but these are restricted to fixed peer-to-peer (P2P) networks between RSUs.

\begin{figure}[!t]
    \centering
    \includegraphics[width=\linewidth]{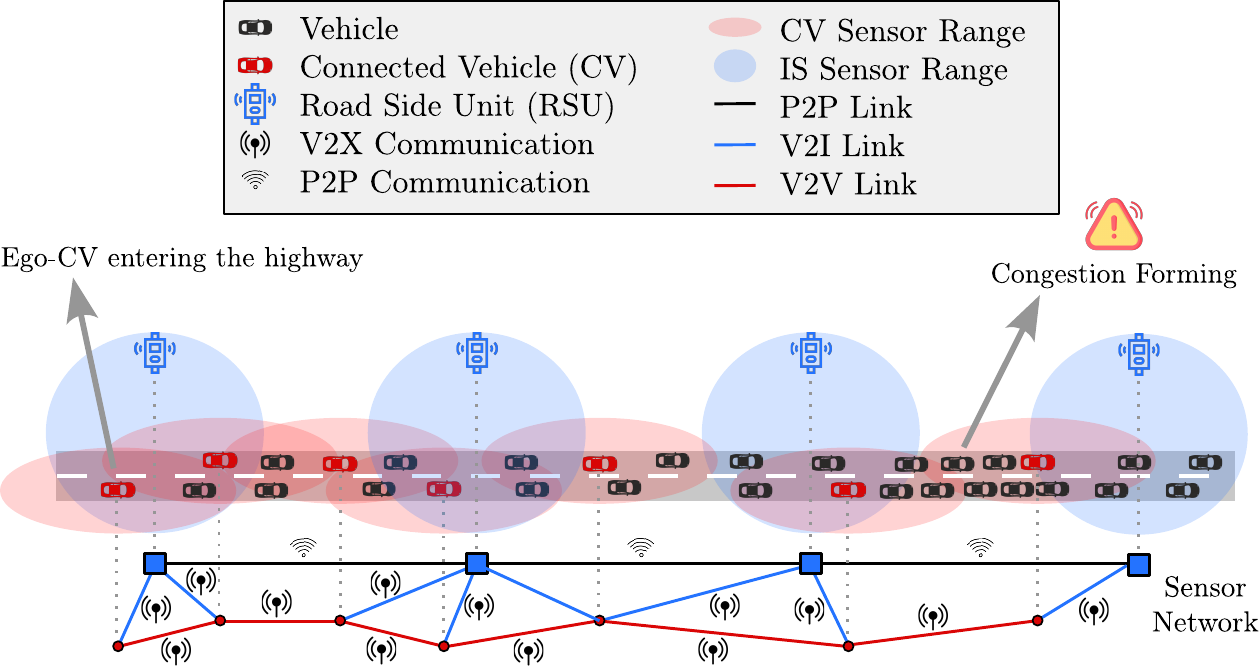}
    \caption{An example scenario of a $2.7 \text{-km}$ two-lane highway with a mix of connected and conventional vehicles.}
    \label{fig:scenario}
\end{figure}

This paper proposes a novel distributed traffic state estimation (DTSE) framework, motivated by the scenario illustrated in Fig.~\ref{fig:scenario}, in which we consider a $2.7 \text{-km}$ two-lane highway with a mix of connected and conventional vehicles. 
Downstream on the highway (to the right), congestion is forming that could lead to a backward-propagating shockwave. 
An ego-CV is entering the highway upstream (on the left) and cannot detect downstream congestion, which is beyond its onboard sensors' range. 
Critically, if this ego-CV could acquire a global traffic state estimate, it would be able to detect the approaching congestion shockwave in advance. 
This foreknowledge would allow the ego-CV to execute a preemptive, non-emergency deceleration maneuver. 
Such an action serves two functions.
First, it prevents the ego-CV from contributing to the congestion's severity upon impact.
Second, from a traffic engineering perspective, it acts as a control input to actively dissipate the congestion wave \citep{lee2025}, thereby enhancing overall string stability and improving the macroscopic traffic throughput of the entire highway segment.

Our main contributions in this paper are as follows. 
We first propose a decentralized estimation architecture where both fixed RSUs and mobile CVs act as active sensing and estimation nodes, exchanging estimates only with their V2X neighbors rather than a distant central server in the TMC. 
Building upon this, we introduce a rigorous distributed Kalman filter (DKF) algorithm specifically designed to leverage the ARZ traffic flow model. 
The proposed framework accurately fuses heterogeneous measurements to reconstruct complex non-equilibrium dynamics, such as traffic shock waves. 
Finally, we provide a comprehensive, scenario-driven evaluation to demonstrate the framework's practical utility. 
We show the ability of our DTSE algorithm to reconstruct the complete spatiotemporal density and relative flow states during a congestion shockwave, even under the real-world conditions of low CV penetration and sparsely placed RSUs.

The remainder of the paper is organized as follows. Section~\ref{sec:traffic-model} introduces the ARZ model and derives its discretized state-space formulation. Section~\ref{sec:sensor-network} describes the sensor network model for both RSUs and CVs. Our DTSE algorithm is presented in Section~\ref{sec:DTSE-algorithm}. Section~\ref{sec:experiments} discusses the experimental results. Finally, Section~\ref{sec:conclusion} provides the conclusion and highlights future research directions.

\section{Traffic Model} \label{sec:traffic-model}
This section reviews the ARZ model for describing highway traffic dynamics. Then, following \cite{vishnoi2024}, we derive its discretized state-space formulation.

\subsection{ARZ Model} \label{subsec:ARZ}
The ARZ model \citep{aw2000,zhang2002} is a second-order macroscopic traffic flow model given by
\begin{align}
    \partial_t \rho + \partial_d (\rho v) &= 0  \label{eq:ARZpde1}\\
    \partial_t \psi +\partial_d \psi v &= -\frac{\rho(v-V_e(\rho))}{\tau} \label{eq:ARZpde2}
\end{align}
where $\rho(t,d)$ is the density (vehicles per unit road length) and $v(t, d)$ is the average speed (unit length per unit time), over time $t$ and spatial position $d$.  Let 
\begin{equation} \label{eq:driver-characteristic}
\chi \coloneqq v+p(\rho)
\end{equation}
denote the driver characteristic with $p(\rho)$ the pressure function that represents anticipatory driving behavior. Then, in~\eqref{eq:ARZpde2}, the auxiliary variable $\psi = \rho \chi$ denotes the relative flow, which is a momentum-like quantity that reflects deviations from the equilibrium traffic conditions, and  $\tau > 0$ is the relaxation time that determines how quickly drivers adapt their speed toward equilibrium. 

The pressure function
$
p(\rho) = v_f \left(\rho / \rho_m \right)^\gamma
$,
where $v_f$ is the free-flow speed, $\rho_m$ is the maximum jam density, and $\gamma>0$ is a fundamental diagram parameter.
Then, the equilibrium velocity at density $\rho$ is given by
$
V_e(\rho) = v_f \left(1-\left(\rho / \rho_m \right)^\gamma\right)
$.
The parameters $v_f, \rho_m, \gamma$ can be calibrated using empirical data and the fundamental diagram.

\subsection{Discretized State-Space Model}
Following \cite{vishnoi2024}, we discretize the ARZ model in space and time. The highway is partitioned into $N$ segments/cells of uniform length $\Delta h$, and the temporal domain is discretized using a sampling interval $\Delta t$. This results in a discretized state-space model with $\rho$ and $\psi$ as finite state vectors representing the average density and relative flow within the discretized cells. The discretized ARZ model can be written in state-space form as
\begin{align}\label{eq:ssARZ}
     x_{k+1} = Ax_{k} + Gf(x_k,u_k) + \omega_k
\end{align}
where $x_k = [x_{1,k}^\T,\ldots,x_{N,k}^\T]^\T\in \mathbb R^{2N}$ denotes the state vector of the highway discretized into $N$ cells, with %the state of cell~$i$
$$
x_{i,k} = \begin{bmatrix}
 \rho_{i,k} &
 \psi_{i,k}  & 
\end{bmatrix}^\T, \quad i=1,\ldots,N
$$
where $\rho_{i,k}$ is the average traffic density and $\psi_{i,k}$ the average relative flow within cell $i$ at time $k$. The variable $\omega_k$ denotes the process noise.

The input $u_k \in \mathbb{R}^3$ in \eqref{eq:ssARZ} specifies the boundary conditions of the model and is given by
\begin{equation}\label{eq:input}
   u_{k} = \begin{bmatrix}
   D_{0,k} & \chi_{0,k} & \rho_{{N+1},k} 
   \end{bmatrix}^\T
\end{equation}
where $D_{0,k}$ is the upstream demand, i.e., the number of vehicles per unit time entering the highway at time step $k$, $\chi_{0,k}$ is the driver characteristic \eqref{eq:driver-characteristic} at the upstream boundary, and $\rho_{N+1,k}$ specifies the density at the downstream boundary that governs the outflow of traffic.

In standard TSE formulations, the input $u_k$ is typically assumed to be known at a centralized TMC. However, in the distributed setting considered here, only the boundary RSUs can directly measure the boundary conditions that determine $u_k$. 
% The boundary RSUs exchange this information with one another via their fixed P2P links, and subsequently broadcast it to any CVs within their V2X communication range.
In this paper, we assume that boundary RSUs broadcast $u_k$ to all nodes via a low-bandwidth, long-range communication channel (e.g., cellular LTE or LoRaWAN). 
This hierarchical communication architecture is well-established for ensuring scalability in heterogeneous vehicular networks \citep{abboud2016, zadobrischi2024}.
Since $u_k$ consists of only three scalar values, the bandwidth overhead for this global broadcast is negligible compared to the dense covariance matrices exchanged locally via short-range V2X.
In a practical deployment, this assumption can be relaxed by including the boundary variables in the consensus vector. Since the boundary RSUs can directly measure $u_k$, they act as leader nodes for these variables, allowing the true values to diffuse rapidly across the network via a consensus protocol \citep{ren2008}.

The system matrices $A,G \in \R^{2N\times 2N}$ in \eqref{eq:ssARZ} are given by
\[
A= I_N \otimes
\begin{bmatrix}
    1 & 0\\
    \frac{v_f}{\tau} & 1-\frac{1}{\tau}
\end{bmatrix},
\quad G = I_N \otimes
\begin{bmatrix}
    \frac{\Delta t}{\Delta h} & 0\\
    0 & \frac{\Delta t}{\Delta h}
\end{bmatrix}
\]
where $\otimes$ denotes the Kronecker product, and 
\begin{equation*}
f(x_k,u_k)=
\biggl[
\begin{bmatrix}
q_{0,k}-q_{1,k}\\
\phi_{0,k}-\phi_{1,k}
\end{bmatrix}^\T,\ldots,
\begin{bmatrix}
q_{N-1,k}-q_{N,k}\\
\phi_{N-1,k}-\phi_{N,k}
\end{bmatrix}^\T
\biggr]^\T
\end{equation*}
represents net flows between adjacent cells. 
Here, $q_{i,k}$ denotes the traffic flux leaving cell $i$ and entering cell $i+1$ at time step $k$, measured in vehicles per unit time. Since there are no on-ramps or off-ramps in Fig.~\ref{fig:scenario}, the flux $q_{i,k}$ across the cell boundary is determined by the minimum of the upstream demand $D_{i,k}$ and the downstream supply $S_{i+1,k}$, i.e.,
$
q_{i,k} = \min (D_{i,k}, S_{{i+1},k}).
$
The variable $\phi_{i,k}$ denotes the relative flux given by
\[
\phi_{i,k} = q_{i,k} \chi_{i,k} = q_{i,k}\dfrac{\psi_{i,k}}{\rho_{i,k}}.
\]
The demand $D_{i,k}$ describes the maximum flow that can exit cell $i$ and is given by
\begin{equation*}
D_{i,k} =
\begin{cases}
    \rho_{i,k}(\chi_{i,k}-p(\rho_{i,k})), &\text{if} \ \rho_{i,k} \leq \sigma (\chi_{i,k}) \\
    \sigma(\chi_{i,k})(\chi_{i,k}-p(\sigma(\chi_{i,k}))), &\text{if} \ \rho_{i,k} > \sigma(\chi_{i,k}) 
\end{cases}    
\end{equation*}
where the critical density $\sigma(\chi_{i,k})$ is given by
\[
\sigma(\chi_{i,k}) = \rho_m\para{\frac{\chi_{i,k}}{v_f(1+\gamma)}}^{1 / \gamma}.
\]
The supply $S_{i,k}$ describes the maximum flow that can enter cell $i$ from upstream. Unlike demand, which depends only on the local state, the supply also accounts for the upstream driver characteristic $\chi_{i-1,k}$, as upstream traffic conditions constrain admissible inflow. The supply
\begin{equation*}
S_{i,k} \!=\!
\begin{cases}
    \sigma(\chi_{i,k})(\chi_{i-1,k}-p(\sigma(\chi_{i-i,k}))), \! & \!\!\! \text{if } \rho_{i,k} \!\leq\! \sigma(\chi_{{i-1},k}) \\
    \rho_{i,k}(\chi_{{i-1},k}-p(\rho_{i,k})), \! & \!\!\! \text{if } \rho_{i,k} \!>\! \sigma(\chi_{{i-1},k}).
\end{cases}
\end{equation*}
As with demand, the critical density $\sigma$ separates the free-flow and congested regimes. When $\rho_{i,k}$ is below the upstream critical value $\sigma(\chi_{i-1,k})$, the downstream cell can accept additional inflow. When the density exceeds this threshold, congestion restricts the admissible inflow.

\section{Sensor Network Model} \label{sec:sensor-network}

Both fixed RSUs and CVs are assumed to measure the density $\rho_{i,k}$ and relative flow $\psi_{i,k}$ of the cell they occupy in the highway. Fixed RSUs always observe a predetermined segment, while CVs provide measurements only for the segment they currently occupy.
CVs measure their individual speed directly using GPS or odometry. 
They can also utilize on-board perception systems, such as camera-based vehicle counting and spacing estimation, to infer a local traffic density $\rho_{i,k}$ in cell~$i$. 
The average speed $v_{i,k}$ of cell~$i$ can be approximated by tracking and aggregating the individual speeds of all vehicles currently in that cell. 
The relative flow $\psi_{i,k}$ cannot be directly observed, but it can be reconstructed from the CVs' local measurements by substituting the measured $\rho_{i,k}$ and the approximated $v_{i,k}$ into the model: $\psi_{i,k} = \rho_{i,k} \chi_{i,k} = \rho_{i,k} (v_{i,k} + p(\rho_{i,k}))$, where $\chi_{i,k}$ is the driver characteristic in \eqref{eq:driver-characteristic} and $p(\rho_{i,k})$ is the anticipatory pressure term described in Section~\ref{subsec:ARZ}.

The measurement equation for sensor $l\in \mathcal S$, where $ \mathcal{S}$ is the set of all sensors (both RSUs and CVs), is given by
\begin{equation}\label{eq:spatmeas}
y^l_k = C_k^l x_k + \nu^l_k
\end{equation}
where $y^l_k \in \mathbb{R}^2$ is the measurement vector collected by sensor $l$ at time $k$, $x_k \in \mathbb{R}^{2N}$ is the global traffic state vector at time $k$ evolving according to \eqref{eq:ssARZ}, $\nu^l_k \in \mathbb{R}^2$ is the measurement noise associated with sensor $l$. 
The dimension of $y^l_k$ is $\mathbb{R}^2$ because each sensor~$l$ measures the two state components, $\rho_{i_{l,k},k}$ and $\psi_{i_{l,k},k}$, associated with cell $i_{l,k}\in\{1,\dots, N\}$ where sensor~$l$ is located at time~$k$.

The measurement matrix $C_k^l \in \mathbb{R}^{2 \times 2N}$ is dynamic and sparse. 
It extracts the state vector block $x_{i_{l, k}, k}$ from $x_k$, where $i_{l,k}$ denotes the index of the specific cell currently occupied by sensor $l$ at time $k$. 
It is constructed as
$$C_k^l = e_{i_{l,k}}^\top \otimes I_2$$
where $i_{l,k} \in \{1, \dots, N\}$ is the cell index occupied by sensor $l$ at time $k$  and $e_{i_{l,k}}$ denotes the $i_{l,k}$-th canonical basis vector of $\mathbb{R}^N$.
This ensures that the measurement $y^l_k$ is directly mapped to the two-component state vector ($\rho_{i_{l,k},k}$ and $\psi_{i_{l,k},k}$) of the cell it occupies at time $k$. 
As RSUs are fixed, the corresponding $i_{l, k}=i_l$ is constant with respect to time $k$. 
For CVs, $i_{l, k}$ is updated based on the vehicle's position at time $k$.

As shown in Fig.~\ref{fig:scenario}, each RSU is connected to its immediate RSU neighbors via a P2P network. In addition to these P2P links, both CVs and RSUs establish further short-range communication links with other CVs within their respective communication ranges. This ensures that the external input $u_k$ can propagate from the boundary RSUs to other RSUs and CVs, which is essential for our proposed DTSE algorithm.

For designing and analyzing the information flow in our DTSE algorithm, the sensor network (comprising both fixed RSUs and mobile CVs) is modeled as a dynamic undirected graph $\mathcal{G}_k = (\mathcal{S}_k, \mathcal{E}_k)$ at each time step $k$. 
Here, $\mathcal{S}_k\subseteq \mathcal{S}$ is the set of all sensor nodes active at time $k$, and an edge $(l, m) \in \mathcal{E}_k$ exists if sensor nodes $l$ and $ m$ are within the physical V2X communication range ($\approx 300-500$ m) of each other, establishing a bidirectional communication link for state estimate exchange. 
Moreover, we adopt the simplifying technical assumption that the volume of transmitted data (the local state estimates and covariance matrices) does not exceed the capacity of any link \citep{olfati-saber2007}. 
In this paper, this assumption is justified by the relatively low instantaneous CV penetration rate expected in the considered operational scenario (illustrated in Fig.~\ref{fig:scenario}). This keeps the node density and, consequently, the number of simultaneous communication sessions low, preventing the channel congestion and packet loss typically observed in dense vehicular ad-hoc networks \citep{kenney2011, abboud2016}. 
This allows us to neglect link-saturation effects and focus solely on the connectivity topology necessary for the distributed estimation consensus.

\section{Distributed Traffic State Estimation Algorithm} \label{sec:DTSE-algorithm}

Estimating the state of a dynamical system via a sensor network is a well-established problem in control theory. 
Foundational studies by \cite{olfati-saber2005, olfati-saber2007} introduced distributed Kalman filtering that employs consensus protocols, enabling sensor nodes to estimate the state by sharing measurements and covariances. 
Key refinements include stability analyses under communication constraints \citep{carli2008}, diffusion strategies for localized information propagation \citep{cattivelli2010}, and optimization of convergence rate under weak observability \citep{das2016}. 
Further advances address event-triggered communication \citep{battistelli2018}, integrated estimation and control \citep{talebi2019}, and robustness to model uncertainty \citep{zorzi2019}.

However, these works focused on linear systems. Although suitable for many sensor networks, standard DKF frameworks do not handle the nonlinear dynamics and strict physical constraints (such as non-negativity and saturation) that are crucial for macroscopic traffic flow applications. 
The algorithm presented below addresses this gap by adopting the information-form DKF \citep{battistelli2018} and extending it to the nonlinear ARZ traffic model by incorporating both linearization and the enforcement of physical constraints on the state estimate.

\subsection{Algorithm}

Let the process noise $\omega_k \sim \mathcal{N}(0, Q)$ with the covariance matrix $Q \in \mathbb{R}^{2N \times 2N}$. Moreover, let the local measurement noise $\nu^l_k \sim \mathcal{N}(0, R^l_k)$ with covariance $R^l_k \in \mathbb{R}^{2 \times 2}$, for $l\in\mathcal S$. At time $k=0$, each sensor node~$l \in \mathcal{S}$ is initialized as
\begin{align*}
	\hat{x}_{0|0}^l = \mathbb{E}[x_0], \quad
	\Xi_{0|0}^l = (P_0)^{-1}, \quad
	\xi_{0|0}^l = \Xi_{0|0}^l \hat{x}_{0|0}^l
\end{align*}
where $\hat{x}_{0|0}^l$ is the initial estimate of sensor~$l$, $\Xi_{0|0}^l$ is the initial information matrix with $P_0$ the initial error covariance, and $\xi_{0|0}^l$ is the initial information vector.

Then, for each time $k \geq 1$ and for each sensor node $l \in \mathcal{S}$, we perform the following five steps:

\subsubsection*{Step 1 - Linearization:}
We linearize \eqref{eq:ssARZ} around the current local posterior estimate $\hat{x}_{k|k}^l$ as
\begin{subequations}
\label{eq:linearized_dyn}
\begin{align}
\Lambda^l_k &= A + G \left. \frac{\partial f(x, u_k)}{\partial x} \right|_{x = \hat{x}_{k|k}^l} 
\label{eq:linearized_dyn-Lambda} \\
\eta^l_k &= G \left( f(\hat{x}_{k|k}^l, u_k) - \left[ \left. \frac{\partial f(x, u_k)}{\partial x} \right|_{x = \hat{x}_{k|k}^l} \right] \hat{x}_{k|k}^l \right)
\label{eq:linearized_dyn-eta}
\end{align}
\end{subequations}
where $\Lambda^l_k$ is the linearized state transition matrix and $\eta^l_k$ is the offset vector.

\subsubsection*{Step 2 - Local Measurement Update:}
Each node assimilates its own sensor data into the information space. 
The local information contribution $\theta_k^l$ and associated precision matrix $\Theta_k^l$ are computed as
$$
\theta_k^l = (C_k^l)^\T (R_k^l)^{-1} y_k^l, \quad \Theta_k^l = (C_k^l)^\T (R_k^l)^{-1} C_k^l .
$$
Then, the local posterior updates are given by
$$
\xi_{k|k}^l = \xi_{k|k-1}^l + \theta_k^l, \quad \Xi_{k|k}^l = \Xi_{k|k-1}^l + \Theta_k^l.
$$

\subsubsection*{Step 3 - Information Fusion:}
Nodes exchange their local information pairs $(\xi_{k|k}^l, \Xi_{k|k}^l)$ with neighbors $j \in \mathcal{N}^l_k$. 
An iterative consensus step is performed for $L$ iterations. 
Let $\alpha = 0, \dots, L$ be the consensus iteration index. Initialize $\xi^{l,(0)}_{k|k} = \xi^l_{k|k}$ and $\Xi^{l,(0)}_{k|k} = \Xi^l_{k|k}$.
Then, at iteration $\alpha=1, \dots, L$, compute
\begin{align*}
\xi^{l,(\alpha)}_{k|k} &= \sum_{j \in \mathcal{N}^l_k \cup \{l\}} \pi_{(l,j),k}  \xi^{j,(\alpha-1)}_{k|k} \\
\Xi^{l,(\alpha)}_{k|k} &= \sum_{j \in \mathcal{N}^l_k \cup \{l\}} \pi_{(l,j),k}  \Xi^{j,(\alpha-1)}_{k|k}
\end{align*}
where $\pi_{(l,j),k}$ are the weights of a doubly stochastic matrix compatible with the communication graph $\mathcal{G}_k$.
After $L$ iterations, the fused information variables are given by
\begin{equation}
\label{eq:fused}
\overline{\xi}^l_{k|k} = \xi^{l,(L)}_{k|k}, \quad \overline{\Xi}^l_{k|k} = \Xi^{l,(L)}_{k|k}.
\end{equation}

\subsubsection*{Step 4 - Prediction:}
Using the linearized dynamics \eqref{eq:linearized_dyn} and the fused information \eqref{eq:fused}, we compute the prior for the next time step. We employ the matrix inversion lemma to propagate the information matrix without direct inversion of the state covariance, i.e.,
\begin{align}
\Xi^l_{k+1|k} = Q^{-1} - Q^{-1} \Lambda^l_k M_k^l (\Lambda^l_k)^\T Q^{-1}
\end{align}
where $M_k^l = \left( \overline{\Xi}^l_{k|k} + (\Lambda^l_k)^\T Q^{-1} \Lambda^l_k \right)^{-1}$.
The predicted information vector accounts for $\eta^l_k$ in \eqref{eq:linearized_dyn-eta} as
$$\xi^l_{k+1|k} = \Xi^l_{k+1|k} \left( \Lambda^l_k (\overline{\Xi}^l_{k|k})^{-1} \overline{\xi}^l_{k|k} + \eta^l_k \right).$$

\subsubsection*{Step 5 - Enforcing Physical Constraints:}
To ensure the traffic state estimate remains physically meaningful, we project the posterior $\hat{x}^l_{k+1} = (\Xi^l_{k+1|k})^{-1} \xi^l_{k+1|k}$ onto the feasible state space \citep{vishnoi2024}. 
For each cell $i \in \{1, \dots, N\}$, $\hat{x}^l_{i,k+1} = [\hat{\rho}^l_{i, k+1} ~ \hat{\psi}^l_{i, k+1}]^\T$ is projected as
\begin{subequations}
\label{eq:projection}
\begin{align}
\hat{\rho}^l_{i, k+1} &\leftarrow \min \Big( \max(\hat{\rho}^l_{i, k+1}, 0), \rho_m \Big) \\
\hat{\psi}^l_{i, k+1} &\leftarrow \min \Big( \max(\hat{\psi}^l_{i, k+1}, 0), v_f \rho_m \Big).
\end{align}
\end{subequations}
Thus, the posterior estimate is updated as
$$\hat{x}^l_{k+1} \leftarrow \Pi((\Xi^l_{k+1|k})^{-1} \xi^l_{k+1|k})$$ where $\Pi$ is the projection operator defined in \eqref{eq:projection}. 
The information matrix $\Xi_{k+1|k}^l$ remains unchanged to preserve uncertainty characteristics, but the information vector is updated to reflect the constrained state as
$$\xi^l_{k+1|k} \leftarrow \Xi^l_{k+1|k} \, \hat{x}^l_{k+1}.$$

\section{Experiment Results} \label{sec:experiments}

\subsection{Experimental Setup}
We validate the proposed algorithm using microscopic traffic data generated via the SUMO simulator. 
The scenario consists of a two-lane highway with a total length of $2.7$ km. 
To align with the macroscopic boundary control formulation, the first and last $100$ m serve as buffer zones for extracting upstream demand and downstream density boundary conditions, yielding an effective study domain of $2.5$ km. 
The simulation runs for a total duration of $1200$ s with a maximum speed limit of $100$ km/h. 
Traffic demand is generated at the upstream boundary using an exponential arrival process with a mean interarrival time of $1$ s, corresponding to a flow of approximately $3600$ veh/h. 
Vehicle dynamics are governed by the standard Kraus car-following model with a minimum gap of $1.5$ m.
To evaluate the estimator's performance under transient, non-equilibrium conditions, we induce a congestion wave using a temporary bottleneck. 
A variable speed limit is applied at the $d=2200$ m mark (relative to the effective domain), reducing the speed limit from $100$ km/h to $10$ km/h during the interval $t \in [700, 760]$ s. This intervention successfully triggers a stop-and-go shockwave that propagates upstream.

The sampling interval is set to $\Delta t = 1$ s. 
The parameters of the ARZ model are calibrated to match the microscopic simulation data as follows: free-flow speed $v_f = 100$ km/h, maximum jam density $\rho_m = 250$ veh/km, and the fundamental diagram parameter $\gamma = 1.25$. 
The relaxation time is set to $\tau = 1$ s; while this value is lower than typical empirical observations, it was found to yield the best fit for the aggressive acceleration dynamics inherent to the default SUMO behavior.
For the implementation of the DTSE algorithm, the highway is discretized into $N=25$ cells of length $\Delta h=100$ m. The stability of discretization is guaranteed since $v_f {\Delta t / \Delta h} = 100 (1000/3600) \times 1/100 \approx 0.28 < 1$, i.e., the Courant-Friedrichs-Lewy condition \citep{courant1967} is satisfied. 

The fixed infrastructure consists of four RSUs positioned at $d \in \{50, 850, 1650, 2450\}$ (m). 
These units maintain persistent P2P links with their immediate neighbors. 
For mobile sensing, we assume a CV penetration rate of approximately $10\%$. 
Both RSUs and CVs operate with a V2X communication range of $400$ m.
The analysis focuses on a specific time window $t \in [700, 842]$ (s), defined by the traversal of a reference ego-CV that enters the network as the congestion wave is triggered. 
During this interval, a total of 253 vehicles traverse the highway segment, 25 of which are designated as CVs.
The DTSE filter is tuned and initialized with the following parameters for all sensors. The initial guess of $\rho_0=\rho_f$ and $\psi_0=v_f\rho_f$, where the free-flow density is $\rho_f = 50$ veh/km, and we take $P_0 = I_{2N\times 2N}$. The measurement noise covariance is set to $R^l = \diag(4, 400)$ for all sensors $l \in \mathcal{S}$.
To account for discretization errors and model mismatch, we utilize a process noise covariance $Q = I_N \otimes \diag(4, 400)$. 
To ensure physical realizability, the state estimates are projected onto the box constraints $\rho \in [0, 250]$ (veh/km) and $\psi \in [0, 25000]$ (veh/h).
Finally, the consensus protocol utilizes $L=5$ communication rounds per time step to accelerate information fusion among the sensor nodes.

\begin{figure*}[!t]
    \centering
    \includegraphics[width=\linewidth]{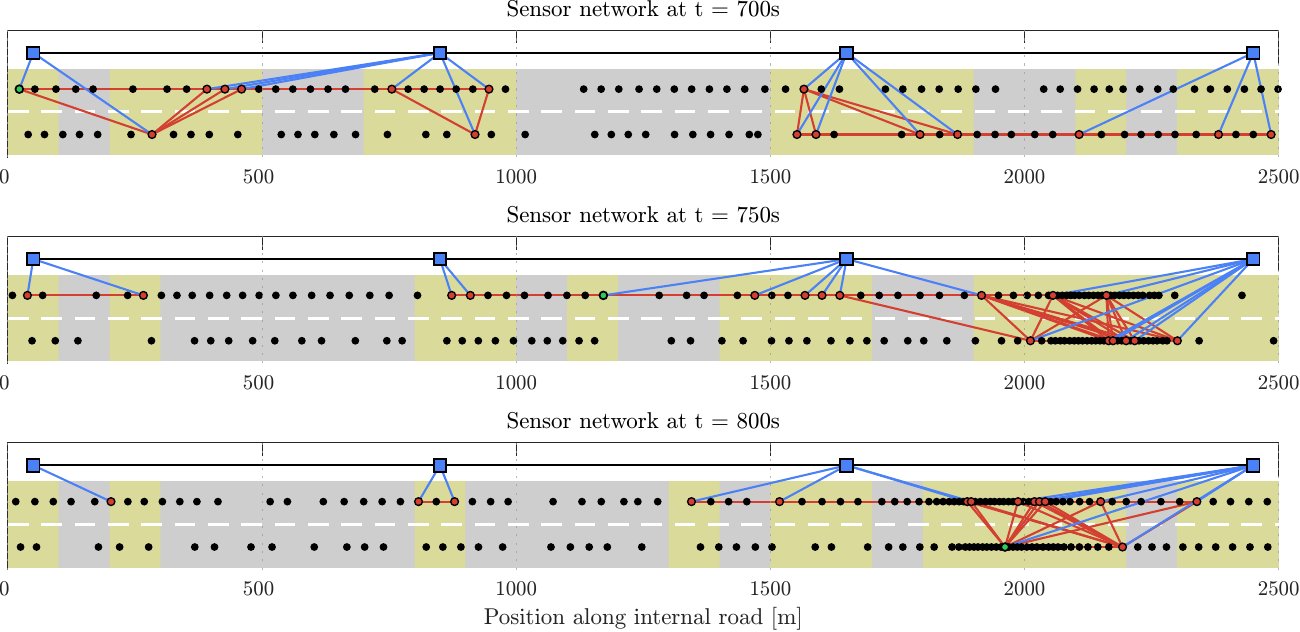}
    \caption{Evolution of the vehicle-sensor communication network at three representative time instants ($t=700$ s, $750$ s, $800$ s. Blue squares denote RSU, red dots represent CVs, black dots indicate vehicles, and the green dot marks the ego-CV. Yellow shaded regions highlight each sensor's local sensing range. The black links indicate P2P communication, red links V2V communication, and blue links V2I communication at each time instants. These panels illustrate how the ego-CV's accessible information changes dynamically as traffic evolves and as communication links form and break due to vehicle motion.}
    \label{fig:FIG2}
\end{figure*}

\begin{figure}
    \centering
    \includegraphics[width=\linewidth]{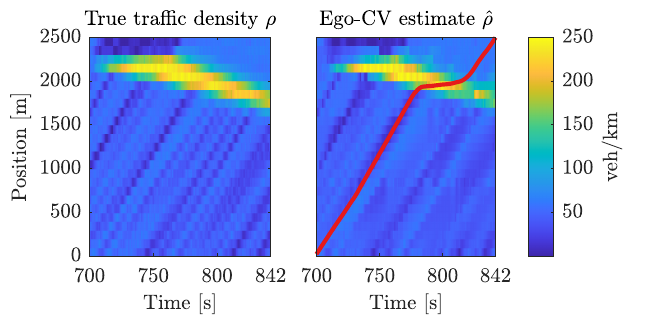}
    \caption{Spatiotemporal reconstruction of highway density by the ego-CV. The left column shows the SUMO density field, and the right column shows the ego-CV estimate. The red line is the trajectory of the ego-CV.}
    \label{fig:FIG1}
\end{figure}

\subsection{Estimation Results}

We evaluate the performance of the proposed DTSE algorithm by analyzing the local estimate maintained by a single representative ego-CV as it traverses the network. This analysis assesses the ability of an individual node (i.e., ego-CV) to reconstruct the global traffic state despite possessing only local sensing capabilities and intermittent connectivity.

Figure~\ref{fig:FIG2} provides a sequence of snapshots illustrating the dynamic evolution of the sensing and communication graph $\mathcal{G}_k$. 
The distinct node types are color-coded: fixed RSUs (blue squares), participating CVs (red dots), and non-connected vehicles (black dots). 
The ego-CV, whose internal estimation belief is the subject of this analysis, is highlighted in green. 
The yellow shaded regions delineate the instantaneous observable subspace of the network, i.e., areas where density and flow are directly measurable by at least one sensor node. 
As evidenced by the gaps between yellow regions, the network is characterized by sparse and fragmented observability. 
Consequently, accurate estimation relies heavily on the distributed consensus mechanism to diffuse information from observed segments to unobserved ones via V2X communication links.

A comparative analysis of the ground truth traffic dynamics and the distributed reconstruction is presented in Fig.~\ref{fig:FIG1}. 
The left panel visualizes the macroscopic density field $\rho(t,d)$ derived directly from the microscopic SUMO simulation, serving as the ground truth. 
The right panel displays the estimated density field $\hat{\rho}(t,d)$ as reconstructed by the ego-CV. To contextualize the estimation perspective, the trajectory of the ego-CV is overlaid on the estimated heatmap.

Qualitatively, the estimator exhibits high fidelity in capturing the fundamental nonlinear features of the traffic flow. 
As shown in the ground truth panel (Fig.~\ref{fig:FIG1}, left), the imposition of the speed limit drop at $t=700$ s triggers a rapid transition from free-flow to congested conditions, resulting in a backward-propagating shockwave (the high-density region slanting upwards to the left).
Despite the ego-CV being located upstream of the bottleneck when the congestion initiates, its local estimate (Fig.~\ref{fig:FIG1}, right) successfully captures the onset of the congestion wave almost simultaneously with the ground truth. 
This indicates effective information dissemination from the downstream RSUs and CVs (located at the bottleneck) upstream to the ego-CV through the multi-hop consensus network. 
Furthermore, the estimator correctly reproduces the propagation speed of the shockwave front and the subsequent dissipation of the jam after the speed limit is restored at $t=760$ s.

A consequence of the diffusion-based consensus and the discretization of the state space is that the distributed estimate naturally exhibits slight smoothing effects compared to the granular ground truth. 
Therefore, such smoothing accurately resolves the spatiotemporal extent of the high-density clusters. 
This result confirms that the proposed nonlinear information filter enables individual nodes to maintain global situational awareness even when the network topology is dynamic and direct observation coverage is partial.

\subsection{Statistical Evaluation}

To complement the specific ego-CV case study, we perform a rigorous statistical evaluation to quantify how the density and spatial distribution of CVs affect global estimation accuracy. 
Since the communication graph topology $\mathcal{G}_k$ and the observability of the traffic state depend heavily on the specific positions of the participating vehicles, a Monte Carlo simulation is necessary to marginalize the effects of random spatial configurations.

We evaluate the estimator's performance across a range of CV penetration rates, denoted by the set $\mathcal{P}=\{2, 5, 10, 15, 20\}$ (\%). 
For each rate $p \in \mathcal{P}$, we conduct $N_\mathrm{trial} = 100$ independent Monte Carlo trials. 
In each trial, a distinct subset of vehicles is randomly designated as CVs from the total pool of 253 unique trajectories traversing the highway during the simulation window.
The reference ego-CV is included in every subset to maintain a consistent estimation node for comparison.

To assess the fidelity of the distributed reconstruction, we employ two complementary metrics evaluated over the time interval $t=1,\dots, N_\mathrm{steps}$ during which the ego-CV is active. 
For each state variable $z \in \{\rho, \psi\}$, we first compute the Root Mean Square Error (RMSE) as
$$\text{RMSE}(z) = \sqrt{\frac{1}{N_\mathrm{trial} N_\mathrm{steps}}\sum^{N_\mathrm{trial}}_{j=1} \sum^{N_\mathrm{steps}}_{k=1} \|z_{k,j} - \hat{z}^\mathrm{ego}_{k,j}\|^2}$$
which quantifies the absolute magnitude of the estimation deviation in physical units (veh/km or veh/h), penalizing large outliers. We also compute the Symmetric Mean Absolute Percentage Error (SMAPE) as
$$\text{SMAPE}(z) = \frac{100}{N_\mathrm{trial} N_\mathrm{steps}} \sum^{N_\mathrm{trial}}_{j=1} \sum^{N_\mathrm{steps}}_{k=1} \frac{2 \| z_{k,j} - \hat{z}^{ego}_{k,j} \|}{\| z_{k,j} \| + \| \hat{z}^\mathrm{ego}_{k,j} \|}.$$
Unlike standard MAPE, which can be numerically unstable near zero density, SMAPE is bounded between 0\% and 200\% and provides a symmetric, scale-independent assessment of relative accuracy. 

\begin{figure}
    \centering
    \includegraphics[width=\linewidth]{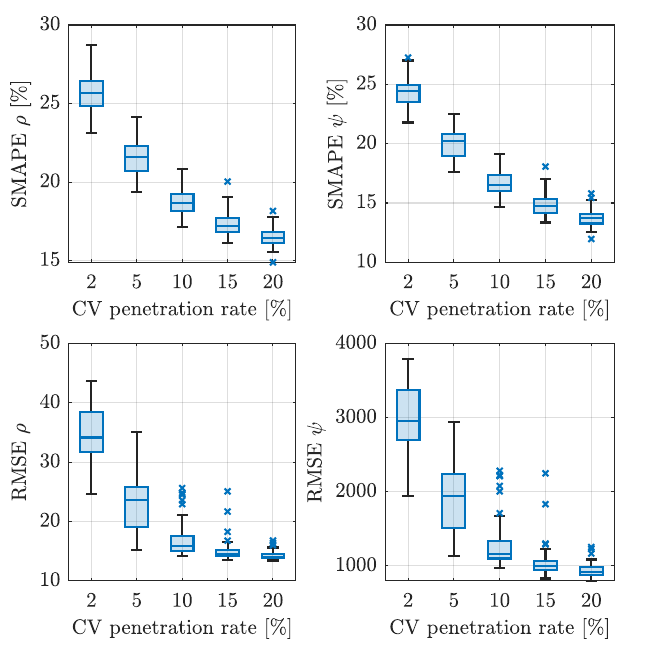}
    \caption{Estimation accuracy of the ego-CV in SMAPE and RMSE for different CV penetration rates over 100 trials.}
    \label{fig:SMAPES}
\end{figure}

Figure~\ref{fig:SMAPES} displays the statistical distribution of these error metrics across the 100 trials for each penetration rate. 
The results reveal a monotonic improvement in estimation fidelity with increasing penetration rate, characterized by two distinct convergence regimes.

At low penetration rates ($2-5\%$), the estimator exhibits not only higher median errors but also significant variance (indicated by the wide interquartile ranges in the box plots). 
In this regime, the performance is highly sensitive to the stochastic spatial arrangement of CVs. 
A lucky distribution may bridge the gaps between RSUs, while an unlucky distribution results in a fragmented network graph $\mathcal{G}_k$. 
When the graph is disconnected, information fusion is halted, preventing the ego-CV from correcting its estimates using data from remote segments. 
The sharp decline in error as the rate approaches $10\%$ suggests the network is passing a percolation threshold, where the probability of forming a giant connected component, linking the ego-CV to the boundary RSUs, increases.

Beyond the $10\%$ threshold, the marginal gain in accuracy diminishes. 
In this regime, the network connectivity is generally robust (high algebraic connectivity), ensuring fast consensus convergence. 
The error reduction here is primarily driven by improved local observability: a higher density of CVs provides a finer spatial resolution of measurements, reducing the discretization and interpolation errors inherent to the macroscopic model. 
The narrowing of the box plot whiskers at $20\%$ confirms that the estimator becomes robust to random variations in CV placement, achieving consistent performance regardless of the specific vehicle subset selected.

\section{Conclusion} \label{sec:conclusion}

This work proposed a distributed traffic state estimation framework that transforms connected vehicles and infrastructure into a cooperative sensor network. This approach addresses the scalability and communication bottlenecks found in centralized estimation. By integrating an information-form distributed Kalman filter with consensus-based protocols, sensor nodes can effectively fuse heterogeneous local data to reconstruct the overall traffic state while maintaining physical flow consistency. Experiments using a microsimulator under non-equilibrium conditions demonstrate that individual nodes can accurately estimate complex spatiotemporal features, like backward-propagating shockwaves, beyond their immediate sensing range. These results show that diffusive information propagation supports reliable state estimation by a connected vehicle, even when the infrastructure sensors are sparsely distributed.

Statistical analysis of the proposed framework reveals a fundamental connection between network topology, connected vehicle penetration rate, and estimation accuracy. Network observability depends not only on the connectivity of the dynamic communication graph, but also on the sensor density. When the penetration rate exceeds a critical threshold, the distributed estimation algorithm achieves asymptotic error convergence and becomes more robust to random variations in the sensor network. Future research will assess the estimator’s performance under imperfect communication channels, validate results with real-world trajectory data, and extend the method to highway networks with on-ramps and off-ramps.

%\begin{ack}
%Place acknowledgments here.
%\end{ack}

%\section*{DECLARATION OF GENERATIVE AI AND AI-ASSISTED TECHNOLOGIES IN THE WRITING PROCESS}
%During the preparation of this work the author(s) used [NAME TOOL / SERVICE] in order to [REASON]. After using this tool/service, the author(s) reviewed and edited the content as needed and take(s) full responsibility for the content of the publication.

%\bibliography{ifacconf} 

\begin{thebibliography}{34}
\providecommand{\natexlab}[1]{#1}
\providecommand{\url}[1]{\texttt{#1}}
\providecommand{\urlprefix}{URL }
\expandafter\ifx\csname urlstyle\endcsname\relax
  \providecommand{\doi}[1]{doi:\discretionary{}{}{}#1}\else
  \providecommand{\doi}{doi:\discretionary{}{}{}\begingroup
  \urlstyle{rm}\Url}\fi

\bibitem[{Abboud et~al.(2016)Abboud, Omar, and Zhuang}]{abboud2016}
Abboud, K., Omar, H.A., and Zhuang, W. (2016).
\newblock Interworking of {DSRC} and cellular network technologies for {V2X}
  communications: A survey.
\newblock \emph{IEEE Transactions on Vehicular Technology}, 65(12), 9457--9470.

\bibitem[{Ansari(2021)}]{ansari2021}
Ansari, K. (2021).
\newblock Joint use of {DSRC} and {C-V2X} for {V2X} communications in the 5.9
  {GHz} {ITS} band.
\newblock \emph{IET Intelligent Transport Systems}, 15(2), 213--224.

\bibitem[{Arthurs et~al.(2021)Arthurs, Gillam, Krause, Wang, Halder, and
  Mouzakitis}]{arthurs2021}
Arthurs, P., Gillam, L., Krause, P., Wang, N., Halder, K., and Mouzakitis, A.
  (2021).
\newblock A taxonomy and survey of edge cloud computing for intelligent
  transportation systems and connected vehicles.
\newblock \emph{IEEE Transactions on Intelligent Transportation Systems},
  23(7), 6206--6221.

\bibitem[{Aw and Rascle(2000)}]{aw2000}
Aw, A. and Rascle, M. (2000).
\newblock Resurrection of ``second order" models of traffic flow.
\newblock \emph{SIAM journal on applied mathematics}, 60(3), 916--938.

\bibitem[{Battistelli et~al.(2018)Battistelli, Chisci, and
  Selvi}]{battistelli2018}
Battistelli, G., Chisci, L., and Selvi, D. (2018).
\newblock A distributed {Kalman} filter with event-triggered communication and
  guaranteed stability.
\newblock \emph{Automatica}, 93, 75--82.

\bibitem[{Bekiaris-Liberis et~al.(2016)Bekiaris-Liberis, Roncoli, and
  Papageorgiou}]{bekiaris2016}
Bekiaris-Liberis, N., Roncoli, C., and Papageorgiou, M. (2016).
\newblock Highway traffic state estimation with mixed connected and
  conventional vehicles.
\newblock \emph{IEEE Transactions on Intelligent Transportation Systems},
  17(12), 3484--3497.

\bibitem[{Carli et~al.(2008)Carli, Chiuso, Schenato, and Zampieri}]{carli2008}
Carli, R., Chiuso, A., Schenato, L., and Zampieri, S. (2008).
\newblock Distributed {Kalman} filtering based on consensus strategies.
\newblock \emph{IEEE Journal on Selected Areas in Communications}, 26(4),
  622--633.

\bibitem[{Cattivelli and Sayed(2010)}]{cattivelli2010}
Cattivelli, F.S. and Sayed, A.H. (2010).
\newblock Diffusion strategies for distributed {Kalman} filtering and
  smoothing.
\newblock \emph{IEEE Transactions on Automatic Control}, 55(9), 2069--2084.

\bibitem[{Courant et~al.(1967)Courant, Friedrichs, and Lewy}]{courant1967}
Courant, R., Friedrichs, K., and Lewy, H. (1967).
\newblock On the partial difference equations of mathematical physics.
\newblock \emph{IBM Journal of Research and Development}, 11(2), 215--234.

\bibitem[{Daganzo(1994)}]{daganzo1994}
Daganzo, C.F. (1994).
\newblock The cell transmission model: A dynamic representation of highway
  traffic consistent with the hydrodynamic theory.
\newblock \emph{Transportation Research Part B: Methodological}, 28(4),
  269--287.

\bibitem[{Das and Moura(2016)}]{das2016}
Das, S. and Moura, J.M. (2016).
\newblock Consensus+innovations distributed {Kalman} filter with optimized
  gains.
\newblock \emph{IEEE Transactions on Signal Processing}, 65(2), 467--481.

\bibitem[{Fountoulakis et~al.(2017)Fountoulakis, Bekiaris-Liberis, Roncoli,
  Papamichail, and Papageorgiou}]{fountoulakis2017}
Fountoulakis, M., Bekiaris-Liberis, N., Roncoli, C., Papamichail, I., and
  Papageorgiou, M. (2017).
\newblock Highway traffic state estimation with mixed connected and
  conventional vehicles: Microscopic simulation-based testing.
\newblock \emph{Transportation Research Part C: Emerging Technologies}, 78,
  13--33.

\bibitem[{Guo et~al.(2017)Guo, Chen, Li, and Zhang}]{guo2017}
Guo, Y., Chen, Y., Li, W., and Zhang, C. (2017).
\newblock Distributed state-observer-based traffic density estimation of urban
  freeway network.
\newblock In \emph{IEEE Conference on Intelligent Transportation Systems},
  1177--1182.

\bibitem[{Herrera and Bayen(2010)}]{Herrera2010}
Herrera, J.C. and Bayen, A.M. (2010).
\newblock Incorporation of lagrangian measurements in freeway traffic state
  estimation.
\newblock \emph{Transportation Research Part B: Methodological}, 44(4),
  460--481.

\bibitem[{Kenney(2011)}]{kenney2011}
Kenney, J.B. (2011).
\newblock Dedicated short-range communications {(DSRC)} standards in the
  {United States}.
\newblock \emph{Proceedings of the IEEE}, 99(7), 1162--1182.

\bibitem[{Lee et~al.(2025)Lee, Wang, Jang, Lichtl{\'e}, Hayat, Bunting,
  Alanqary, Barbour, Fu, Gong et~al.}]{lee2025}
Lee, J.W., Wang, H., Jang, K., Lichtl{\'e}, N., Hayat, A., Bunting, M.,
  Alanqary, A., Barbour, W., Fu, Z., Gong, X., et~al. (2025).
\newblock Traffic control via connected and automated vehicles {(CAVs)}: An
  open-road field experiment with 100 {CAVs}.
\newblock \emph{IEEE Control Systems}, 45(1), 28--60.

\bibitem[{Lighthill and Whitham(1955)}]{lighthill1955}
Lighthill, M.J. and Whitham, G.B. (1955).
\newblock On kinematic waves {II.} a theory of traffic flow on long crowded
  roads.
\newblock \emph{Proceedings of the Royal Society of London. Series A.
  Mathematical and Physical Sciences}, 229(1178), 317--345.

\bibitem[{Mendes et~al.(2025)Mendes, Araujo, Goes, Corujo, and
  Oliveira}]{mendes2025}
Mendes, B., Araujo, M., Goes, A., Corujo, D., and Oliveira, A.S. (2025).
\newblock Exploring {V2X} in {5G} networks: A comprehensive survey of
  location-based services in hybrid scenarios.
\newblock \emph{Vehicular Communications}, 100878.

\bibitem[{Olfati-Saber(2005)}]{olfati-saber2005}
Olfati-Saber, R. (2005).
\newblock Distributed {Kalman} filter with embedded consensus filters.
\newblock In \emph{IEEE Conference on Decision and Control}, 8179--8184.

\bibitem[{Olfati-Saber(2007)}]{olfati-saber2007}
Olfati-Saber, R. (2007).
\newblock Distributed {Kalman} filtering for sensor networks.
\newblock In \emph{IEEE Conference on Decision and Control}, 5492--5498.

\bibitem[{Ren and Beard(2008)}]{ren2008}
Ren, W. and Beard, R.W. (2008).
\newblock \emph{Distributed Consensus in Multi-Vehicle Cooperative Control:
  Theory and Applications}.
\newblock Springer.

\bibitem[{Richards(1956)}]{richards1956}
Richards, P.I. (1956).
\newblock Shock waves on the highway.
\newblock \emph{Operations Research}, 4(1), 42--51.

\bibitem[{Seo et~al.(2017)Seo, Bayen, Kusakabe, and Asakura}]{seo2017review}
Seo, T., Bayen, A.M., Kusakabe, T., and Asakura, Y. (2017).
\newblock Traffic state estimation on highway: A comprehensive survey.
\newblock \emph{Annual reviews in control}, 43, 128--151.

\bibitem[{Sun and Work(2014)}]{sun2014}
Sun, Y. and Work, D.B. (2014).
\newblock A distributed local kalman consensus filter for traffic estimation.
\newblock In \emph{IEEE Conference on Decision and Control}, 6484--6491.

\bibitem[{Sun and Work(2017)}]{sun2017}
Sun, Y. and Work, D.B. (2017).
\newblock Scaling the {Kalman} filter for large-scale traffic estimation.
\newblock \emph{IEEE Transactions on Control of Network Systems}, 5(3),
  968--980.

\bibitem[{Talebi and Werner(2019)}]{talebi2019}
Talebi, S.P. and Werner, S. (2019).
\newblock Distributed {Kalman} filtering and control through embedded average
  consensus information fusion.
\newblock \emph{IEEE Transactions on Automatic Control}, 64(10), 4396--4403.

\bibitem[{Van~Hinsbergen et~al.(2011)Van~Hinsbergen, Schreiter, Zuurbier,
  Van~Lint, and Van~Zuylen}]{hinsbergen2012}
Van~Hinsbergen, C.P.I.J., Schreiter, T., Zuurbier, F.S., Van~Lint, J.W.C., and
  Van~Zuylen, H.J. (2011).
\newblock Localized extended kalman filter for scalable real-time traffic state
  estimation.
\newblock \emph{IEEE Transactions on intelligent transportation systems},
  13(1), 385--394.

\bibitem[{Vishnoi et~al.(2024)Vishnoi, Nugroho, Taha, and
  Claudel}]{vishnoi2024}
Vishnoi, S.C., Nugroho, S.A., Taha, A.F., and Claudel, C.G. (2024).
\newblock Traffic state estimation for connected vehicles using the
  second-order {Aw-Rascle-Zhang} traffic model.
\newblock \emph{IEEE Transactions on Intelligent Transportation Systems},
  25(11), 16719--16733.

\bibitem[{Vivas et~al.(2015)Vivas, Siri, Ferrara, Sacone, Cavanna, and
  Rubio}]{vivas2015}
Vivas, C., Siri, S., Ferrara, A., Sacone, S., Cavanna, G., and Rubio, F.R.
  (2015).
\newblock Distributed consensus-based switched observers for freeway traffic
  density estimation.
\newblock In \emph{IEEE Conference on Decision and Control}, 3445--3450.

\bibitem[{Yuan et~al.(2012)Yuan, Van~Lint, Wilson, van Wageningen-Kessels, and
  Hoogendoorn}]{yuan2012}
Yuan, Y., Van~Lint, J., Wilson, R.E., van Wageningen-Kessels, F., and
  Hoogendoorn, S.P. (2012).
\newblock Real-time {Lagrangian} traffic state estimator for freeways.
\newblock \emph{IEEE Transactions on Intelligent Transportation Systems},
  13(1), 59--70.

\bibitem[{Zadobrischi and Havriliuc(2024)}]{zadobrischi2024}
Zadobrischi, E. and Havriliuc, S. (2024).
\newblock Enhancing scalability of {C-V2X} and {DSRC} vehicular communication
  protocols with {LoRa} 2.4 {GHz} in the scenario of urban traffic systems.
\newblock \emph{Electronics}, 13(14), 2845.

\bibitem[{Zhang(2002)}]{zhang2002}
Zhang, H. (2002).
\newblock A non-equilibrium traffic model devoid of gas-like behavior.
\newblock \emph{Transportation Research Part B: Methodological}, 36(3),
  275--290.

\bibitem[{Zhang and Lu(2020)}]{zhang2020}
Zhang, L. and Lu, Y. (2020).
\newblock Distributed consensus-based boundary observers for freeway traffic
  estimation with sensor networks.
\newblock In \emph{American Control Conference}, 4497--4502.

\bibitem[{Zorzi(2019)}]{zorzi2019}
Zorzi, M. (2019).
\newblock Distributed {Kalman} filtering under model uncertainty.
\newblock \emph{IEEE Transactions on Control of Network Systems}, 7(2),
  990--1001.

\end{thebibliography}

\end{document}